\documentclass[published]{epl}
\usepackage{amsmath, amssymb}
\usepackage{graphicx}
\usepackage{subfigure}
\newcommand{\bb}{\begin{equation}}
\newcommand{\en}{\end{equation}}

\title{Distribution of counterions near discretely charged planes and rods}
\shorttitle{Counterions at discretely charged surfaces}
\author{M. L. Henle\inst{1} \and C. D. Santangelo\inst{1} \and D. M. Patel\inst{1} \and P. A. Pincus\inst{1,2,3}}
\institute{
\inst{1} Department of Physics, University of California, Santa Barbara, California 93106\\
\inst{2} Department of Materials, University of California, Santa Barbara, California 93106\\
\inst{3} Program in Biomolecular Science and Engineering, University of California, Santa Barbara, California 93106\\}
\shortauthor{M.L. Henle \etal}

\issue{66}{2}{2004}{284}{15 April 2004}
\recff{24 October 2003}{10 February 2004}
\begin{document}

\maketitle

\begin{abstract}
Realistic charged macromolecules are characterized by discrete (rather than homogeneous) charge distributions.  We investigate the effects of surface charge discretization on the counterion distribution at the level of mean-field theory using a two-state model.   Both planar and cylindrical geometries are considered; for the latter case, we compare our results to  numerical solutions of the full Poisson-Boltzmann equation.  We find that the discretization of the surface charge can cause enhanced localization of the counterions near the surface; for charged cylinders, counterion condensation can exceed  Oosawa-Manning condensation. 
\end{abstract}

\section{Introduction}

The interaction between charged macromolecules and counterions is a crucial component of the physics of charged systems~\cite{safran}.  Although realistic macromolecules are often composed of discrete charges, theoretical models typically assume a homogeneous surface charge distribution.  In the case of cylindrical polyelectrolytes (PEs), this assumption leads to Oosawa-Manning (OM) condensation~\cite{manning, oosawa}, in which counterions become closely associated with the macromolecule, effectively lowering its overall charge.  A similar effect also occurs for spherical macromolecules~\cite{alexander}.   Theoretically, the effects of surface charge inhomogeneities on the counterion distribution has been studied for planar~\cite{moreira, lukatsky}, cylindrical~\cite{abascal, zimm1, guldbrand, chang}, and spherical~\cite{messina, allahyarov} geometries.  In particular, both Moreira \emph{et al.}~\cite{moreira} and Lukatsky \emph{et al.}~\cite{lukatsky} have shown analytically that the heterogeneity of a planar charge distribution leads to an enhanced localization of the counterions near the macromolecular surface.  Thus, theoretical models that assume homogeneous surface charge distributions -- including OM condensation theory -- must be modified for realistic charge distributions. 

In this letter, we explore the effects of surface charge discretization on the counterion distribution in the absence of added salt.  For small charge modulation, it is appropriate to use a perturbative expansion of the Poisson-Boltzmann (PB) equation to describe the mean-field behavior of the counterion distribution.  It has already been shown that such an expansion leads to enhanced counterion localization for planes~\cite{lukatsky}, and it is straightforward to show that the same conclusion holds for surfaces of arbitrary geometry~\cite{santangelo}.  For a discrete surface charge distribution, however, pair association between discrete macromolecular charges and the discrete counterions~\cite{stell, fisher} can become important; that is, a large number of counterions can become localized near the  macromolecular surface. In PB theory, the counterions are treated as a smooth distribution function, rather than as discrete objects.  In calculating the electrostatic energy, such a description clearly fails when many counterions are spatially localized, because it integrates over these localized regions.  As a result, the (coarse-grained) self-energy of the localized counterions is incorrectly included in the electrostatic energy.  Thus, PB theory cannot sufficiently describe the counterion distribution of a discretely charged surface in this regime. 

Motivated by these considerations, we introduce a two-state model (TSM) which explicitly allows for the possibility of surface charge-counterion pair association.  Counterions in the bound state are confined to the surface by the electrostatic attraction to a single discrete surface charge; free counterions, on the other hand, ``see'' the macromolecule and the bound counterions as a homogeneous charge distribution.  We ignore any correlations in the thermal fluctuations of the counterions, but explicitly take into account the discreteness of the bound counterions.  Our main result is that, for both planar and cylindrical geometries, the TSM predicts that surface charge discretization leads to enhanced counterion localization.  In particular, it is possible to have counterion condensation on PEs that exceeds the amount predicted by OM condensation theory.  We also find that, for cylindrical geometries, the TSM yields excellent agreement with numerical solutions to the PB equation when the number of bound counterions is small, but that the PB equation dramatically under-predicts the extent of counterion localization when this number is large.

\section{Two-state model}  

For a macromolecule with an arbitrary surface charge distribution, the mean-field counterion density profile is determined by the competition of the electrostatic attraction to the macromolecule with both the counterion entropy and the interaction between counterions. For a discrete surface charge distribution, the localized counterions will experience a strong electrostatic attraction to a discrete surface charge.  In order to capture the effects of surface charge discretization on the counterion distribution, we allow the counterions to exist in one of two possible states:  in the bound state, counterions bind to a single charge on the surface, sacrificing most of their entropy to gain electrostatic energy; in the free state, by contrast, counterions gain less electrostatic energy but have a much larger entropy.  Thus, the two-state model is essentially a picture for counterion association with, and the consequent charge regulation of, a discretely charge surface.  The possibility of counterion association for homogeneously charged objects has been considered by several other authors~\cite{levin, lau, shklovskii}. 

Consider an aqueous solution of charged macromolecules and their neutralizing counterions.  We use the cell model, in which each macromolecule and its neutralizing counterions (of valency $Q$) are enclosed in a  cell of some volume $V$; interactions between cells are ignored.  We also work exclusively within the primitive model, where the aqueous solvent is approximated as a homogeneous medium with a constant dielectric constant $\epsilon$ ($\epsilon \approx 80$ for water).  The surface $\Sigma_s$ of each macromolecule has a total area $A_s$ and is decorated with a lattice (with lattice constant $d$) of $N_s$ discrete surface charges of valency $Q_s$.   Let  $N_f$ be the number of counterions in the free state and $N_b$ the number of counterions in the bound state; by charge neutrality, $Q_s N_s = Q N_f+ Q N_b$. The bound counterions lie on a surface $\Sigma_b$ (of total area $A_b$), which is of the same geometry as  $\Sigma_s$ but separated from it by $\delta$, the distance of closest approach (set, for example, by the size of the counterions).    Because of electrostatic repulsion and excluded volume interactions, we only allow one counterion per binding site.  Defining the free counterion density  $n_f (\vec{x}) =  \sum_{i=1}^{N_f} \delta^3(\vec{x}-\vec{r}_i)$,  and assuming $\delta \ll d$, the electrostatic energy can be written as
\begin{eqnarray}
\label{eq:2stateH}
& & \beta {\cal H} =  l_B Q \iint\limits_{V} \frac{\upd^3 x \, \upd^3 x ' \,}{\left|\vec{x}-\vec{x}'\right|} n_f (\vec{x}) \left[\frac{Q}{2}  n_f (\vec{x}' )+Q N_b n_b (\vec{x}' )-Q_s N_s n_s (\vec{x}' ) \right] -\frac{l_B Q Q_s N_b}{\delta}\\
\notag
& &
-  l_B Q Q_s N_b N_s  \iint\limits_{V-R_1}\frac{\upd^3 x \, \upd^3 x ' \,}{\left|\vec{x}-\vec{x}'\right|} n_b (\vec{x}) n_s (\vec{x}' )+\frac{l_B Q^2 N_b^2}{2} \iint\limits_{V-R_2} \frac{\upd^3 x \, \upd^3 x ' \,}{\left|\vec{x}-\vec{x}'\right|} n_b (\vec{x}) n_b (\vec{x}') 
\end{eqnarray}
with $l_B =  e^2/ \epsilon k_B T$ being the Bjerrum length ($l_B = 7.1$ \AA\ in water),  $e$ the elementary unit of charge, $k_B$ Boltzmann's constant, $T$ the temperature, and $\beta = 1/k_B T$.   $n_s (\vec{x})$ and $n_b (\vec{x})$ are the distribution functions for the surface and bound charges, respectively.  Note that the dependence of $\beta {\cal H}$ on $N_s$ and $N_b$ has been written out explicitly, so that the normalizations of  $n_s(\vec{x})$ and $n_b(\vec{x})$ are set to unity.  

In order to estimate the terms involving the surface charges and bound counterions in eq.~(\ref{eq:2stateH}), we make a series of approximations. First, the bound counterions are treated at the level of mean-field theory.  This is equivalent to distributing the charge of the bound counterions equally among each binding site.  In addition,  the discrete nature of the surface charges and the bound counterions can be ignored when calculating most of the terms in eq.~(\ref{eq:2stateH}).  In the limit $\delta \ll d$, charge discreteness clearly must be taken into account when calculating the interaction of a surface charge with the counterion charge in its binding site; we have therefore written this term out explicitly in the electrostatic energy [the second term in eq.~(\ref{eq:2stateH})].  We can approximate the other terms in eq.~(\ref{eq:2stateH}), however, by smearing out the surface charges and the bound counterions on $\Sigma_s$ and $\Sigma_b$, respectively.  Consequently, we set $n_s (\vec{x})=1/A_s$ on $\Sigma_s$ and $0$ elsewhere; similarly, we set $n_b (\vec{x})=1/A_b$ on $\Sigma_b$ and $0$ elsewhere.  Since the interactions of every surface charge with the counterion charge in its binding site are accounted for in the second term of eq.~(\ref{eq:2stateH}), we must be careful not to also include the coarse-grained approximation to this interaction.  Therefore, we exclude the appropriate region $R_1$ from the second integral in eq.~(\ref{eq:2stateH}), so that $|\vec{x}-\vec{x}'|$ does not run over the regions in which this interaction is being calculated.  Finally, the last integral in eq.~(\ref{eq:2stateH}) is an approximation to the sum $l_B Q^2 N_b^2/(2 N_s) \sum_{i \neq j}^{N_s} 1/| \vec{b}_i-\vec{b}_j |$, where $\vec{b}_i$ are the binding site locations.  Because this sum excludes the $i=j$ terms, the integral should exclude the region $R_2$ in which $|\vec{x}-\vec{x}'|$ lies within the Wigner-Seitz cell of a single binding site.

The free energy  $\beta F = \beta {\cal H} - S_f - S_b$, where $S_f = - \int \upd^3 x \, n_f(\vec{x})  (\ln [n_f(\vec{x}) v_0]-1)$  and $S_b = \ln [N_s!/N_b! (N_s-N_b)!]-N_b \ln [v_0/ v_b]$ are the entropies of the free and bound counterions, respectively, with $v_0$ being the volume of a counterion and $v_b$ the volume within which each bound counterion is confined.  The two-state counterion distribution is found by minimizing the grand free energy, $ \beta G=\beta F-\mu [N_b + \int n_f(\vec{x}) \upd^3x]$ with respect to $n_f (\vec{x})$ and $N_b$ (the chemical potential $\mu$ acts as a Lagrange multiplier which enforces charge neutrality).  Minimization of $G$ with respect to $n_f(\vec{x})$ yields the expected Boltzmann distribution, $n_f(\vec{x}) = n_0 \exp\left[-y(\vec{x})\right]$, with $n_0 \equiv e^{\mu}/v_0$.  The potential $y(\vec{x})$ obeys the Poisson-Boltzmann (PB) equation, $\vec{\nabla}^2 y (\vec{x})=-\kappa^2 \exp[-y(\vec{x})]$, where $\kappa^2 = 4 \pi l_B Q^2 n_0$.  Because we have approximated $n_s (\vec{x})$ and $n_b (\vec{x})$ as homogeneous functions in eq.~(\ref{eq:2stateH}), we can use Gauss' Law to show that the boundary condition for $y(\vec{x})$ at $\Sigma_b$ is identical to the boundary condition for a  ``renormalized'' charge $Q N_b - Q_s N_s$ distributed uniformly across the surface $\Sigma_b$.  Thus, the free counterion charge distribution in the TSM is given by the solution to the PB equation for the appropriate geometry, but with a renormalized and uniform surface charge distribution. 

The number of bound counterions $N_b$ is found by minimizing the grand free energy with respect to $N_b$,
\begin{eqnarray}
\label{eq:Gprime}
& & - \ln \left(\frac{N_s}{N_b} - 1 \right)+  \int\limits_{V} \upd^3 x \, n_b(\vec{x}) y(\vec{x}) -  l_B Q^2 N_b \iint\limits_{R_2} \frac{\upd^3 x \, \upd^3 x ' \,}{\left|\vec{x}-\vec{x}'\right|} n_b(\vec{x}) n_b(\vec{x}')\\
\notag
& &
 + l_B Q Q_s N_s  \iint\limits_{R_1} \frac{\upd^3 x \, \upd^3 x ' \,}{\left|\vec{x}-\vec{x}'\right|} n_b(\vec{x}) n_s(\vec{x}') -\ln\left(n_0 v_b\right) -\frac{l_B Q Q_s}{\delta} =0.
\end{eqnarray}

The entropy of the bound counterions gives rise to the first term in eq.~(\ref{eq:Gprime}), a term which clearly diverges as $N_b \rightarrow 0$.  Therefore, the solution to eq.~(\ref{eq:Gprime}) must be some value of $N_b>0$; in other words, a discretely charged surface of arbitrary geometry will always have some bound counterions in the two-state model.

In order to quantify the effects of surface charge discretization on the counterion distribution, we can compare the results of the TSM with the PB solution for the corresponding homogeneously charged surface.  To do so, we need the PB potential $y(\vec{x})$ for a homogeneously charged surface of the appropriate geometry.   Analytically, the solution to the PB equation is known only in planar and cylindrical geometries, so we restrict ourselves to these geometries in what follows.  

For planar macromolecules, $\Sigma_s$ is the plane $z=-\delta$, $\Sigma_b$ is the plane $z= 0$, and the enclosing cell is the positive half-space $z>0$.  Consider the fraction $\phi$ of counterions enclosed between $\Sigma_b$ and a plane at $z>0$:  $\phi(z; N_b) = 1-  Q A/(Q_s N_s) \int_z^\infty \upd z' \, n_f (z'; N_b)$, where $A$ is the area of the plane. For the homogeneously charged plane,  the enclosed fraction is given by $\phi (z;N_b=0)$; for a discretely charged plane, the number of bound counterions is determined by eq.~(\ref{eq:Gprime}).  If we assume that the surface charges are in a square lattice with lattice constant $d$, then both $R_1$ and $R_2$ are the region $\{|x-x'|<d, |y-y'|<d\}$.  In the limit $\delta \ll d$, eq.~(\ref{eq:Gprime}) reduces to
\bb
\label{eq:sigmaR}
\left[\frac{1}{\sigma - \sigma_r}-\frac{Q_s}{Q \sigma} \right] \sigma_r^2 \left(1+\sqrt{2}\right)^{- 8 l_B Q d \sigma_r} = \frac{Q_s}{2 \pi l_B Q \sigma v_b} \exp \left(-\frac{l_B Q Q_s }{\delta}\right)
\en
where $\sigma = Q_s N_s/A$ is the bare charge density of the plane and $\sigma_r = (Q_s N_s - Q N_b)/A$ is the renormalized charge density of the plane.   The free counterion density is given by $n_f (z; N_b) = [2 \pi l_B Q^2 \left(z+\zeta(\sigma_r) \right)^2] ^{-1}$, where $\zeta(\sigma) = 1/(2\pi l_B Q \sigma)$ is the Gouy-Chapman length~\cite{safran}. Clearly, $n_f (z; N_b)$ is a monotonically decreasing function of $N_b$ for any $z>0$.   This in turn implies that  $\phi (z; N_b>0) > \phi (z; 0)$ for any $z>0$. Thus, the two state model predicts that charge discretization of a planar macromolecule leads to enhanced counterion localization, in agreement with the results of \cite{lukatsky,moreira}.

For cylindrical macromolecules, we consider the simplest cylindrically symmetric  charge distribution, a linear array of charges separated by a distance $d$ (the results are qualitatively unchanged for a charged cylinder with a non-zero radius).  In this case, $\Sigma_s$ is the line $z=0$, $\Sigma_b$ is the surface of a cylinder of radius $\delta$, the enclosing cell is a cylinder of radius $R$, and both $R_1$ and $R_2$ are the region $|z-z'|<d$.   $\phi$ is defined to be the fraction of counterions enclosed between $\Sigma_b$ and a cylinder of radius $\delta<r<R$:  $\phi(r; N_b) = 1- 2 \pi L Q/(Q_s N_s) \int_{r}^R \upd r' \, n_f (r'; N_b)$, where $L$ is the length of the cylinder.  Again, the enclosed fraction is given by $\phi (r;N_b=0)$ for the homogeneously charged cylinder; for the discretely charged cylinder, eq.~(\ref{eq:Gprime}) becomes, in the limit $\delta \ll d$,
\bb
\label{eq:xiR}
\left[ \frac{1}{\xi-\xi_r} -\frac{Q_s}{Q \xi}\right] \left(\gamma_r^2 +(1-\xi_r)^2\right) \left(\frac{\delta}{2 d}\right)^{2\xi_r} = \frac{2 \pi d \delta^2}{v_b} \exp \left(-\frac{l_B Q Q_s }{\delta}\right) \equiv \alpha
\en
where $\xi = l_b Q Q_s/d$ is the bare Manning parameter, $\xi_r = l_B Q (Q_s N_s - Q N_b)/L$ is the renormalized Manning parameter , $\xi_0 \equiv \rho/(1+\rho)$, $\rho \equiv \ln[R/\delta]$, and the integration constant $\gamma_r$ is given by the implicit equation $\xi_r=(1+\gamma_r^2)/(1+\gamma_r \cot \left[ \gamma_r \rho \right] )$, with $0<\gamma_r<\pi/\rho$. Eq.~(\ref{eq:xiR}) applies for $\xi_r>\xi_0$; for $\xi_r<\xi_0$, $\gamma_r \rightarrow i \tilde{\gamma}_r$, with $0<\tilde{\gamma}_r<1$.   The crossover $\xi_r =\xi_0$ occurs when $\alpha = \alpha_0 \equiv (1+\rho)^{-2} [(\xi-\xi_0)^{-1}-Q_s/Q \xi ] (\delta/2 d )^{2 \xi_0}$; for $\alpha<\alpha_0$, $\xi_r<\xi_0$, and vice-versa.  Note that, since eq.~(\ref{eq:xiR}) always gives $\xi_r < \xi$, this crossover can occur only when $\xi>\xi_0$.  For cylindrical geometries, the free counterion density is given by \cite{alfrey, katchalsky}
\bb
 \label{eq:PBcyl}
 n_f (r; N_b)=
 \frac{\gamma_r^2 \left( 1+\gamma_r^2 \right) \left( 1+\cot^2\left[\gamma_r \ln \left(\frac{r}{R} \right)\right] \right)}{2 \pi l_B Q^2 r^2 \left( 1-\gamma_r \cot \left[ \gamma_r \ln \left(\frac{r}{R} \right) \right] \right)^2 }
\en
for $\xi_r>\xi_0$ (again, for $\xi_r<\xi_0$, $\gamma_r \rightarrow i \tilde{\gamma}_r$). Careful examination of eq.~(\ref{eq:PBcyl}) reveals that, for any $r$, $n_f (r)$ is a monotonically increasing function of $\xi_r$ (i.e. a monotonically decreasing function of $N_b$). In other words, $\phi (r; N_b>0) > \phi (r; 0)$ for any $\delta<r<R$. Thus, as in the case of planar geometries, we conclude that charge discretization of a cylindrical charge distribution leads to enhanced counterion localization near the macromolecular surface.  

Oosawa-Manning (OM) condensation \cite{manning, oosawa} predicts that some of the counterions of a highly charged polyelectrolyte (PE) will ``condense'' near the surface of the PE, renormalizing the charge of the PE and leading to an effective Manning parameter $\xi_{eff} = 1$.  A rigorous definition of OM condensation can be derived from the PB solution for a homogeneously charged cylinder by taking the infinite dilution limit, $\rho \rightarrow \infty$ \cite{zimm2}.  Above the ``Manning threshold" $\xi=1$, a certain fraction $f=1-1/\xi$ of the counterions do not escape to infinity, and are therefore considered to be condensed; below the Manning threshold, there are no condensed counterions.  For the two-state model, this definition of condensation needs to be modified slightly.  In particular, all of the bound counterions in the TSM are clearly condensed. The free counterions, on the other hand, are still treated with the PB solution for a homogeneously charged cylinder; therefore, the results of OM theory still apply to them.  In particular, if $\xi_r>\xi_0$, then a fraction $f = 1- 1/\xi_r$ of the free counterions are ``Manning condensed'' in the TSM, in addition to the bound counterions.  Thus, when $\xi_r>\xi_0$, the total number of condensed counterions in the TSM and in OM theory are in fact equal.  It is important to distinguish between the bound and Manning condensed counterions here:  the former are confined in a small volume near the surface of the PE, whereas the latter are contained within a more diffuse  ``Manning layer'' in the neighborhood of the PE surface \cite{zimm2}.  As is clear from eq.~(\ref{eq:xiR}), it is also possible to have $\xi_r<\xi_0$.  In this case, no additional free counterions are Manning condensed in the TSM, and $\xi_{eff} = \xi_r<1$.  This can occur both when $\xi>1$, where there are some condensed counterions in OM theory, and when $\xi<1$, where OM theory predicts no condensation.  In both cases, the fraction of condensed counterions in the TSM exceeds the prediction of OM theory.  In other words, it is possible to have counterion condensation \emph{beyond} that predicted by OM theory in the two-state model.  It is important to note, however, that this is only possible at finite dilution; in the infinite dilution limit, it is straightforward to show that eq. (\ref{eq:xiR}) implies that the total number of condensed counterions in the TSM always agrees with OM theory.   

\section{Comparison of the TSM with numerics}

To test the validity of the TSM, we have solved the Poisson-Boltzmann equation numerically for a linear array of fixed charges.  The discrete PB equation in this geometry is given by
\bb
\label{eq:PBnumerical}
PB[ y(\vec{x})]=\frac{1}{4 \pi l_B Q} \nabla^2 y(\vec{x}) + \frac{N_s  e^{- y(\vec{x})} \theta (r-\delta)}{\int \upd^3 x e^{- y(\vec{x})} \theta (r-\delta)} - \frac{1}{\pi (\Delta r/2)^2 \Delta z} \sum_{i=1}^{N_s} \delta_{r,0} \delta_{z,z_i} =0
\en
where $\theta(x) =1$ for $x>0$ and $0$ for $x<0$ , and $z_i$ are the locations of the surface charges, which are assumed to be monovalent for simplicity (i.e. $Q_s=1$).  Due the azimuthal symmetry of the problem, the Laplacian $\nabla^2 = \partial_r^2 +\partial_r/r+\partial_z^2$.  The second derivatives are approximated using a symmetric three step finite difference equation.  An explicit, forward time relaxation scheme is used to find the solution to eq.~(\ref{eq:PBnumerical}): $ y^{(t+1)}(\vec{x})= y^{(t)}(\vec{x})-\Delta t PB[ y^{(t)}(\vec{x})]$.  Convergence takes approximately $10^5$ timesteps with $\Delta t = 0.1$.  Physical lengths are determined by setting the lattice spacings $\Delta r = \Delta z = 0.1 l_B$.  The boundary condition at $r=R$ is $\partial y/\partial r =0$.  Periodic boundary conditions are used in the $z$ direction, with enough lattice points in each period to contain $5$ fixed charges. 
\begin{figure} 
\threeimages[scale=0.65]{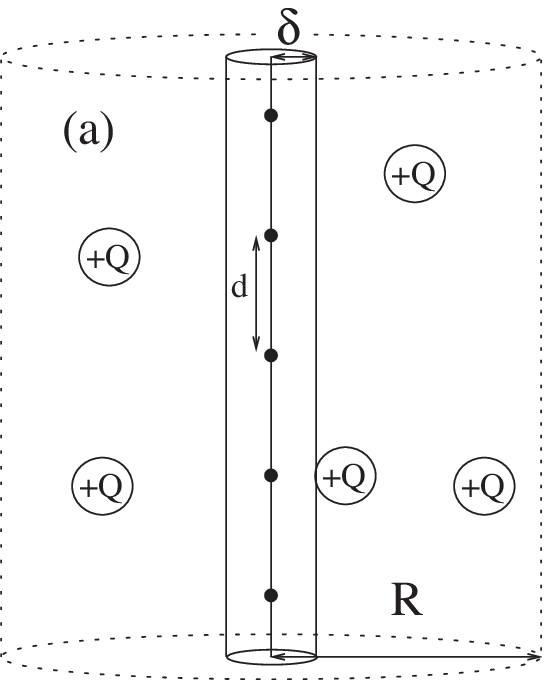}{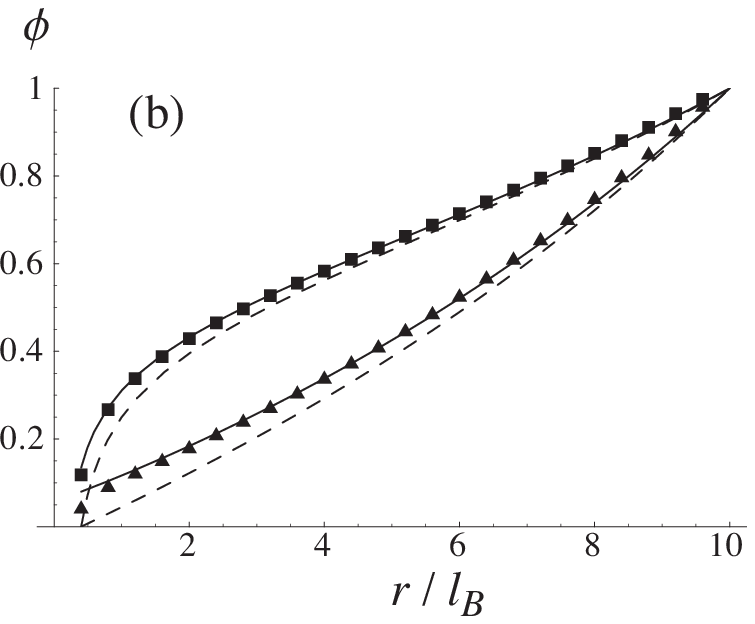}{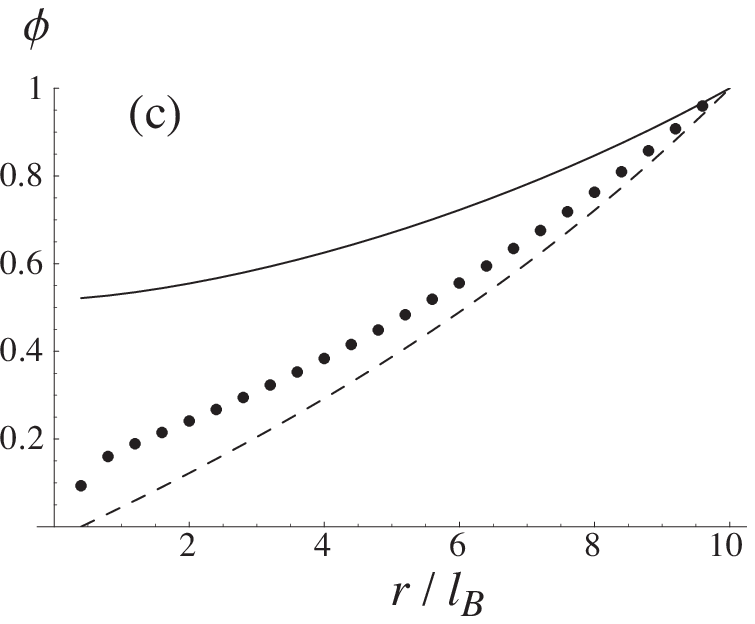}
\caption{a) Schematic representation of a linear array of charges surrounded by neutralizing counterions.  b-c) Fraction of enclosed counterions $\phi (r)$ for $\delta =0.4 l_b, R=10 l_b$, and various values of $Q$ and $d$.  Symbols correspond to the  full PB solutions, solid lines to predictions of the TSM, and dotted lines to the PB solutions for the homogeneously charged cylinder.  b) Results for $Q=3, d= 6 l_b$ (triangles) and $Q=3, d= 2 l_b$ (squares), corresponding to $\xi=0.5$ (below Manning threshold) and $\xi=1.5$ (above Manning threshold), respectively.  The fraction of bound counterions in the TSM is $f_b=0.08$ and $0.13$, respectively.  c) Results for $Q=4, d= 8 l_b$ (circles), corresponding to $\xi=0.5$.  Here, $f_b = 0.52$. }
\label{fig1}
\end{figure}
In fig.~\ref{fig1} we compare the function $\phi (r)$ calculated using the TSM with the numerical solutions to the PB equation.  To illustrate the effects of surface charge discretization, we also show the PB solution for the corresponding homogeneously charged cylinder. $v_b$, which does not need to be specified in the numerics, is set to $4\pi \delta^3/3$ for the TSM.  When the fraction of counterions in the bound state $f_b =1 - \xi_r/\xi$ is small, then the agreement between the TSM and the numerics is excellent (fig.~\ref{fig1}b).  When $f_b$ is large, however, the TSM predicts a much stronger localization of the counterions than is seen in the numerical solutions (fig.~\ref{fig1}c).  This is due to the fact that the PB equation does not include the discrete nature of the counterions; rather, the counterion distribution is a smooth function.  That is, the possibility of surface charge-counterion pair association -- an effect explicitly included in the two-state model -- is excluded from the numerical solutions to the PB equation.  When the binding energy of a surface charge-counterion pair is small  -- in other words, when $Q$ is small or $\delta$ is large -- the number of surface charge-counterion pairs is small, and the PB equation should accurately capture the effects of surface charge discretization on the counterion distribution.  Thus, our numerical results provide verification of the two-state model in this limit.  When the number of surface charge-counterion pairs is large, however, the PB equation is deficient, and we do not expect good agreement between the TSM and our numerical results.   

In conclusion, we have introduced a simple two-state model to describe the counterion distribution for discretely charged macromolecules.  However, both Poisson-Boltzmann theory and strong-coupling (SC) theory~\cite{netz, moreira} can be applied to the same systems as the TSM, and it is important to understand in which regimes each theory is valid.  PB theory is valid when the interaction of discrete surface charges with discrete counterions is sufficiently weak.  SC theory, on the other hand, is valid in precisely the opposite limit, where the interactions between counterions can be treated as a perturbation to the surface charge-counterion interactions.  Finally, in the TSM, interactions between counterions are fully included (albeit at the mean-field level), and the possibility of strong counterion-surface charge interactions is accounted for by allowing discrete counterions to bind to the surface.  Thus, we expect that the TSM is valid in regions between the SC regime and the PB regime, where the interactions of the counterions with both themselves and the discrete surface charges are important.  This can be tested by performing Monte Carlo simulations, which we plan to do in a future work.

\acknowledgments

We thank A. W. C. LAU for useful discussions.  MLH, CDS, and PAP acknowledge the support of the MRL Program of the National Science Foundation under Award No. DMR00-80034 and NSF Grant No. DMR02-037555.   MLH also acknowledges the support of a National Science Foundation Graduate Research Fellowship.  DMP acknowledges the support of NSF Grant No. DMR-0312097 and the use of the UCSB-MRL Central Computing Facilities supported by the NSF.


\begin{thebibliography}{0}

\bibitem{safran}
\Name{Safran S. A.}
\Book{Statistical Thermodynamics of Surfaces, Interfaces, and Membranes}
\Publ{Addison-Wesley Publishing Com., Reading}
\Year{1994}.

\bibitem{manning}
\Name{Manning G. S.}
\REVIEW{Q. Rev. Biophys.}{7}{1969}{179}.

\bibitem{oosawa}
\Name{Oosawa F.}
\Book{Polyelectrolytes}
\Publ{Marcel Decker, New York}
\Year{1971}.


\bibitem{alexander}
\Name{Alexander S., Chaikin P. M., Grant P., Morales G. J. \and Pincus P.}
\REVIEW{J. Chem. Phys.}{80}{1984}{5776}.

\bibitem{moreira}
\Name{Moreira A. G. \and Netz R. R.}
\REVIEW{Europhys. Lett.}{57}{2002}{911}, and references therein.

\bibitem{lukatsky}
\Name{Lukatsky D. B., Safran S. A., Lau A. W. C. \and Pincus P.}
\REVIEW{Europhys. Lett.}{58}{2002}{785}, and references therein;
\Name{Lukatsky D. B. \and Safran S.A.}
\REVIEW{Europhys. Lett.}{60}{2002}{629}.


\bibitem{abascal}
\Name{Gil Montoro J. C. \and Abascal J. L. F.}
\REVIEW{Mol. Phys.}{89}{1996}{1071};
\Name{Gil Montoro J. C. \and Abascal J. L. F.}
\REVIEW{J. Chem. Phys.}{103}{1995}{8273};
\Name{Gil Montoro J. C. \and Abascal J. L. F.}
\REVIEW{J. Chem. Phys.}{109}{1998}{6200}.

\bibitem{zimm1}
\Name{Le Bret M. \and Zimm B. H.}
\REVIEW{Biopolymers}{1984}{23}{271}.

\bibitem{guldbrand}
\Name{Guldbrand L. \and Nordenski\"{o}ld}
\REVIEW{J. Phys. Chem.}{91}{1987}{5714};
\Name{Guldbrand L.}
\REVIEW{Mol. Phys.}{67}{1989}{217}

\bibitem{chang}
\Name{Chang R. \and Yethiraj A.}
\REVIEW{J. Chem. Phys.}{116}{2002}{5284}

\bibitem{messina}
\Name{Messina R., Holm C. \and Kremer K.}
\REVIEW{Eur. Phys. J. E}{4}{2001}{363};
\Name{Messina R}
\REVIEW{Physica A}{308}{2002}{59}.

\bibitem{allahyarov}
\Name{Allahyarov E., L\"{o}wen H. Louis A. A. \and Hansen J. P.}
\REVIEW{Europhys. Lett.}{57}{2001}{731}

\bibitem{santangelo}
\Name{Santangelo C. D.}
unpublished.

\bibitem{stell}
\Name{Stell G. R., Wu K. C. \and Larsen B.}
\REVIEW{Phys. Rev. Lett}{37}{1976}{1369}.

\bibitem{fisher}
\Name{Fisher M. E. \and Levin Y.}
\REVIEW{Phys. Rev. Lett.}{71}{1993}{3826}.

\bibitem{levin}
\Name{Levin Y.}
\REVIEW{Europhys. Lett.}{34}{1996}{405}.

\bibitem{lau}
\Name{Lau A. W. C., Lukatsky D. B., Pincus P. \and Safran S. A.}
\REVIEW{Phys. Rev. E.}{65}{2002}{051502}

\bibitem{shklovskii}
\Name{Shklovskii B. I.}
\REVIEW{Phys. Rev. E.}{60}{1999}{5802}

\bibitem{alfrey}
\Name{Alfrey T., Berg P. W., \and Morawetz H.}
\REVIEW{J. Polym. Sci.}{1951}{7}{153}.

\bibitem{katchalsky}
\Name{Fuoss R. M., Katchalsky A. \and Lifson S.}
\REVIEW{Proc. Natl. Acad. Sci,}{1951}{37}{579}.

\bibitem{zimm2}
\Name{Le Bret M. \and Zimm B. H.}
\REVIEW{Biopolymers}{23}{1987}{287}.

\bibitem{netz}
\Name{Netz R. R.}
\REVIEW{Eur. Phys. J. E}{5}{2001}{557}

\end{thebibliography}
\end{document}